\documentclass[aps,prl,twocolumn,groupedaddress,showpacs,final,superscriptaddress]{revtex4}
\usepackage{graphicx}
\usepackage{amssymb}
\usepackage{stmaryrd}

\begin{document}

\preprint{Version 3.4 Feb 21, 2005}
%%%%%%%%%%%%%%%%%%%%%%%%%%%%%%%%%%%%%%%%%%%%%

\newcommand{\nbb}[0]{{0\nu}\beta\beta}
\newcommand{\fp}[0]{f_\pi}
\newcommand{\eq}[1]{eq.~(\ref{#1})}
\newcommand{\eqs}[2]{eqs.~(\ref{#1,#2})}
\newcommand{\Eq}[1]{Eq.~(\ref{#1})}
\newcommand{\Eqs}[2]{Eqs.~(\ref{#1},\ref{#2})}
\newcommand{\fig}[1]{fig.~\ref{#1}}
\newcommand{\figs}[2]{figs.~\ref{#1},\ref{#2}}
\newcommand{\Fig}[1]{Fig.~\ref{#1}}
\newcommand{\qbb}[0]{Q_{\beta\beta}}
\newcommand{\tl}[0]{\text{L}}
\newcommand{\tr}[0]{\text{R}}
\newcommand{\nc}[0]{N_{\text{c}}}
\newcommand{\mw}[0]{M_{\text{W}}}
\newcommand{\mz}[0]{M_{\text{Z}}}
\newcommand{\mr}[0]{M_{\text{R}}}
\newcommand{\md}[0]{m_{\text{D}}}
\newcommand{\mn}[0]{m_\nu}
\newcommand{\lh}[0]{\Lambda_{\text{H}}}
\newcommand{\aif}[0]{a^{ff^\prime}_{i;ll^\prime}}
\newcommand{\gs}[0]{\Gamma_{\text{S}}=1}
\newcommand{\gps}[0]{\Gamma_{\text{PS}}=\gamma^5}
\newcommand{\lr}[0]{\Lambda_{\text{R}}}
\newcommand{\wrt}[0]{W_{\text{R}}}
\newcommand{\wl}[0]{W_{\text{L}}}
\newcommand{\ls}[0]{\Lambda_{\text{S}}}
\newcommand{\gf}[0]{G_{\text{F}}}
\newcommand{\mm}[0]{M_{\text{M}}}
\newcommand{\sst}[1]{{\scriptscriptstyle #1}}
\newcommand{\beq}{\begin{equation}}
\newcommand{\eeq}{\end{equation}}
\newcommand{\beqa}{\begin{eqnarray}}
\newcommand{\eeqa}{\end{eqnarray}}
\newcommand{\dida}[1]{/ \!\!\! #1}
\renewcommand{\Im}{\mbox{\sl{Im}}}
\renewcommand{\Re}{\mbox{\sl{Re}}}
\def\simge{\hspace*{0.2em}\raisebox{0.5ex}{$>$}
     \hspace{-0.8em}\raisebox{-0.3em}{$\sim$}\hspace*{0.2em}}
\def\simle{\hspace*{0.2em}\raisebox{0.5ex}{$<$}
     \hspace{-0.8em}\raisebox{-0.3em}{$\sim$}\hspace*{0.2em}}
\def\dn{{d_n}}
\def\de{{d_e}}
\def\datom{{d_{\sst{A}}}}
\def\grhobar{{{\bar g}_\rho}}
\def\gpibar{{{\bar g}_\pi^{(I) \prime}}}
\def\gpibaro{{{\bar g}_\pi^{(1) \prime}}}
\def\gpibart{{{\bar g}_\pi^{(2) \prime}}}
\def\mx{{M_X}}
\def\mrho{{m_\rho}}
\def\qpv{{Q_{\sst{W}}}}
\def\lamtv{{\Lambda_{\sst{TVPC}}}}
\def\lamtvs{{\Lambda_{\sst{TVPC}}^2}}
\def\lamtvc{{\Lambda_{\sst{TVPC}}^3}}

%\hspace{-0.8em}\raisebox{-0.3em}{$\sim$}\hspace*{0.2em}}
\def\bra#1{{\langle#1\vert}}
\def\ket#1{{\vert#1\rangle}}
\def\coeff#1#2{{\scriptstyle{#1\over #2}}}
\def\undertext#1{{$\underline{\hbox{#1}}$}}
\def\hcal#1{{\hbox{\cal #1}}}
\def\sst#1{{\scriptscriptstyle #1}}
\def\eexp#1{{\hbox{e}^{#1}}}
\def\rbra#1{{\langle #1 \vert\!\vert}}
\def\rket#1{{\vert\!\vert #1\rangle}}
\def\lsim{{ <\atop\sim}}
\def\gsim{{ >\atop\sim}}
\def\nubar{{\bar\nu}}
\def\psibar{{\bar\psi}}
\def\Gmu{{G_\mu}}
\def\alr{{A_\sst{LR}}}
\def\wpv{{W^\sst{PV}}}
\def\evec{{\vec e}}
\def\notq{{\not\! q}}
\def\notl{{\not\! \ell}}
\def\notk{{\not\! k}}
\def\notp{{\not\! p}}
\def\notpp{{\not\! p'}}
\def\notder{{\not\! \partial}}
\def\notcder{{\not\!\! D}}
\def\notA{{\not\!\! A}}
\def\notv{{\not\!\! v}}
\def\Jem{{J_\mu^{em}}}
\def\Jana{{J_{\mu 5}^{anapole}}}
\def\nue{{\nu_e}}
\def\mns{{m^2_{\sst{N}}}}
\def\me{{m_e}}
\def\mes{{m^2_e}}
\def\mq{{m_q}}
\def\mqs{{m_q^2}}
\def\mw{{M_{\sst{W}}}}
\def\mz{{M_{\sst{Z}}}}
\def\mzs{{M^2_{\sst{Z}}}}
\def\ubar{{\bar u}}
\def\dbar{{\bar d}}
\def\sbar{{\bar s}}
\def\qbar{{\bar q}}
\def\sstw{{\sin^2\theta_{\sst{W}}}}
\def\gv{{g_{\sst{V}}}}
\def\ga{{g_{\sst{A}}}}
\def\pv{{\vec p}}
\def\pvs{{{\vec p}^{\>2}}}
\def\ppv{{{\vec p}^{\>\prime}}}
\def\ppvs{{{\vec p}^{\>\prime\>2}}}
\def\qv{{\vec q}}
\def\qvs{{{\vec q}^{\>2}}}
\def\xv{{\vec x}}
\def\xpv{{{\vec x}^{\>\prime}}}
\def\yv{{\vec y}}
\def\tauv{{\vec\tau}}
\def\sigv{{\vec\sigma}}

\def\sst#1{{\scriptscriptstyle #1}}
\def\gpnn{{g_{\sst{NN}\pi}}}
\def\grnn{{g_{\sst{NN}\rho}}}
\def\gnnm{{g_{\sst{NNM}}}}
\def\hnnm{{h_{\sst{NNM}}}}
\def\xivz{{\xi_\sst{V}^{(0)}}}
\def\xivt{{\xi_\sst{V}^{(3)}}}
\def\xive{{\xi_\sst{V}^{(8)}}}
\def\xiaz{{\xi_\sst{A}^{(0)}}}
\def\xiat{{\xi_\sst{A}^{(3)}}}
\def\xiae{{\xi_\sst{A}^{(8)}}}
\def\xivtez{{\xi_\sst{V}^{T=0}}}
\def\xivteo{{\xi_\sst{V}^{T=1}}}
\def\xiatez{{\xi_\sst{A}^{T=0}}}
\def\xiateo{{\xi_\sst{A}^{T=1}}}
\def\xiva{{\xi_\sst{V,A}}}
\def\rvz{{R_{\sst{V}}^{(0)}}}
\def\rvt{{R_{\sst{V}}^{(3)}}}
\def\rve{{R_{\sst{V}}^{(8)}}}
\def\raz{{R_{\sst{A}}^{(0)}}}
\def\rat{{R_{\sst{A}}^{(3)}}}
\def\rae{{R_{\sst{A}}^{(8)}}}
\def\rvtez{{R_{\sst{V}}^{T=0}}}
\def\rvteo{{R_{\sst{V}}^{T=1}}}
\def\ratez{{R_{\sst{A}}^{T=0}}}
\def\rateo{{R_{\sst{A}}^{T=1}}}
\def\mro{{m_\rho}}
\def\mks{{m_{\sst{K}}^2}}
\def\mpi{{m_\pi}}
\def\mpis{{m_\pi^2}}
\def\mom{{m_\omega}}
\def\mphi{{m_\phi}}
\def\Qhat{{\hat Q}}
\def\FOS{{F_1^{(s)}}}
\def\FTS{{F_2^{(s)}}}
\def\GAS{{G_{\sst{A}}^{(s)}}}
\def\GES{{G_{\sst{E}}^{(s)}}}
\def\GMS{{G_{\sst{M}}^{(s)}}}
\def\GATEZ{{G_{\sst{A}}^{\sst{T}=0}}}
\def\GATEO{{G_{\sst{A}}^{\sst{T}=1}}}
\def\mdax{{M_{\sst{A}}}}
\def\mustr{{\mu_s}}
\def\rsstr{{r^2_s}}
\def\rhostr{{\rho_s}}
\def\GEG{{G_{\sst{E}}^\gamma}}
\def\GEZ{{G_{\sst{E}}^\sst{Z}}}
\def\GMG{{G_{\sst{M}}^\gamma}}
\def\GMZ{{G_{\sst{M}}^\sst{Z}}}
\def\GEn{{G_{\sst{E}}^n}}
\def\GEp{{G_{\sst{E}}^p}}
\def\GMn{{G_{\sst{M}}^n}}
\def\GMp{{G_{\sst{M}}^p}}
\def\GAp{{G_{\sst{A}}^p}}
\def\GAn{{G_{\sst{A}}^n}}
\def\GA{{G_{\sst{A}}}}
\def\GETEZ{{G_{\sst{E}}^{\sst{T}=0}}}
\def\GETEO{{G_{\sst{E}}^{\sst{T}=1}}}
\def\GMTEZ{{G_{\sst{M}}^{\sst{T}=0}}}
\def\GMTEO{{G_{\sst{M}}^{\sst{T}=1}}}
\def\lamd{{\lambda_{\sst{D}}^\sst{V}}}
\def\lamn{{\lambda_n}}
\def\lams{{\lambda_{\sst{E}}^{(s)}}}
\def\bvz{{\beta_{\sst{V}}^0}}
\def\bvo{{\beta_{\sst{V}}^1}}
\def\Gdip{{G_{\sst{D}}^\sst{V}}}
\def\GdipA{{G_{\sst{D}}^\sst{A}}}
\def\fks{{F_{\sst{K}}^{(s)}}}
\def\FIS{{F_i^{(s)}}}
\def\fpi{{F_\pi}}
\def\fk{{F_{\sst{K}}}}
\def\RAp{{R_{\sst{A}}^p}}
\def\RAn{{R_{\sst{A}}^n}}
\def\RVp{{R_{\sst{V}}^p}}
\def\RVn{{R_{\sst{V}}^n}}
\def\rva{{R_{\sst{V,A}}}}
\def\xbb{{x_B}}
\def\mlq{{M_{\sst{LQ}}}}
\def\mlqs{{M_{\sst{LQ}}^2}}
\def\lscal{{\lambda_{\sst{S}}}}
\def\lvect{{\lambda_{\sst{V}}}}
\def\PR#1{{{\em   Phys. Rev.} {\bf #1} }}
\def\PRC#1{{{\em   Phys. Rev.} {\bf C#1} }}
\def\PRD#1{{{\em   Phys. Rev.} {\bf D#1} }}
\def\PRL#1{{{\em   Phys. Rev. Lett.} {\bf #1} }}
\def\NPA#1{{{\em   Nucl. Phys.} {\bf A#1} }}
\def\NPB#1{{{\em   Nucl. Phys.} {\bf B#1} }}
\def\AoP#1{{{\em   Ann. of Phys.} {\bf #1} }}
\def\PRp#1{{{\em   Phys. Reports} {\bf #1} }}
\def\PLB#1{{{\em   Phys. Lett.} {\bf B#1} }}
\def\ZPA#1{{{\em   Z. f\"ur Phys.} {\bf A#1} }}
\def\ZPC#1{{{\em   Z. f\"ur Phys.} {\bf C#1} }}
\def\etal{{{\em   et al.}}}
\def\delalr{{{delta\alr\over\alr}}}
\def\pbar{{\bar{p}}}
\def\lamchi{{\Lambda_\chi}}
\def\qw0{{Q_{\sst{W}}^0}}
\def\qwp{{Q_{\sst{W}}^P}}
\def\qwn{{Q_{\sst{W}}^N}}
\def\qwe{{Q_{\sst{W}}^e}}
\def\qem{{Q_{\sst{EM}}}}
\def\gae{{g_{\sst{A}}^e}}
\def\gve{{g_{\sst{V}}^e}}
\def\gvf{{g_{\sst{V}}^f}}
\def\gaf{{g_{\sst{A}}^f}}
\def\gvu{{g_{\sst{V}}^u}}
\def\gau{{g_{\sst{A}}^u}}
\def\gvd{{g_{\sst{V}}^d}}
\def\gad{{g_{\sst{A}}^d}}
\def\gvftil{{\tilde g_{\sst{V}}^f}}
\def\gaftil{{\tilde g_{\sst{A}}^f}}
\def\gvetil{{\tilde g_{\sst{V}}^e}}
\def\gaetil{{\tilde g_{\sst{A}}^e}}
\def\gvqtil{{\tilde g_{\sst{V}}^e}}
\def\gaqtil{{\tilde g_{\sst{A}}^e}}
\def\gvutil{{\tilde g_{\sst{V}}^e}}
\def\gautil{{\tilde g_{\sst{A}}^e}}
\def\gvdtil{{\tilde g_{\sst{V}}^e}}
\def\gadtil{{\tilde g_{\sst{A}}^e}}
\def\delp{{\delta_P}}
\def\delzp{{\delta_{00}}}
\def\deld{{\delta_\Delta}}
\def\dele{{\delta_e}}
\def\lnew{{{\cal L}_{\sst{NEW}}}}
\def\osffp{{{\cal O}_{7a}^{ff'}}}
\def\oszg{{{\cal O}_{7c}^{Z\gamma}}}
\def\osgg{{{\cal O}_{7b}^{g\gamma}}}
\def\slash#1{#1\!\!\!{/}}
\def\beq{\begin{eqnarray}}
\def\eeq{\end{eqnarray}}
\def\bea{\begin{eqnarray*}}
\def\eea{\end{eqnarray*}}
\def\NCA{\em Nuovo~Cimento}
\def\IJMP{\em Intl.~J.~Mod.~Phys.}
\def\NP{\em Nucl.~Phys.}
\def\PLB{{\em Phys.~Lett.}~B}
\def\JETPLett{{\em JETP Lett.}}
\def\PRL{\em Phys.~Rev.~Lett.}
\def\MPL{\em Mod.~Phys.~Lett.}
\def\PRD{{\em Phys.~Rev.}~D}
\def\PR{\em Phys.~Rev.}
\def\PRP{\em Phys.~Rep.}
\def\ZPC{{\em Z.~Phys.}~C}
\def\PTP{{\em Prog.~Theor.~Phys.}}
% Some other macros used in the sample text
\def\Baryon{{\rm B}}
\def\Lepton{{\rm L}}
\def\sbar{\overline}
\def\stilde{\widetilde}
\def\st{\scriptstyle}
\def\sst{\scriptscriptstyle}
\def\vac{|0\rangle}
\def\argh{{{\rm arg}}}
\def\G{\stilde G}
\def\Wmess{W_{\rm mess}}
\def\NI{\stilde N_1}
\def\antivac{\langle 0|}
\def\infinity{\infty}
\def\mco{\multicolumn}
\def\epp{\epsilon^{\prime}}
\def\psibar{\overline\psi}
\def\nmess{N_5}
\def\chibar{\overline\chi}
\def\lagr{{\cal L}}
\def\drbar{\overline{\rm DR}}
\def\msbar{\overline{\rm MS}}
\def\conj{{{\rm c.c.}}}
\def\Et{{\slashchar{E}_T}}
\def\Etot{{\slashchar{E}}}
\def\mZ{m_Z}
\def\MPlanck{M_{\rm P}}
\def\mW{m_W}
\def\cbeta{c_{\beta}}
\def\sbeta{s_{\beta}}
\def\cW{c_{W}}
\def\sW{s_{W}}
\def\deltaeps{\delta}
\def\sigmabar{\overline\sigma}
\def\epsilonbar{\overline\epsilon}
\def\vep{\varepsilon}
\def\ra{\rightarrow}
\def\half{{1\over 2}}
\def\ko{K^0}
\def\be{\beq}
\def\ee{\eeq}
\def\bea{\begin{eqnarray}}
\def\eea{\end{eqnarray}}
\def\alr{A_{\sst{LR}}}

%  \gsim and \lsim provide >= and <= signs.
\def\centeron#1#2{{\setbox0=\hbox{#1}\setbox1=\hbox{#2}\ifdim
\wd1>\wd0\kern.5\wd1\kern-.5\wd0\fi
\copy0\kern-.5\wd0\kern-.5\wd1\copy1\ifdim\wd0>\wd1
\kern.5\wd0\kern-.5\wd1\fi}}
\def\ltap{\;\centeron{\raise.35ex\hbox{$<$}}{\lower.65ex\hbox{$\sim$}}\;}
\def\gtap{\;\centeron{\raise.35ex\hbox{$>$}}{\lower.65ex\hbox{$\sim$}}\;}
\def\gsim{\mathrel{\gtap}}
\def\lsim{\mathrel{\ltap}}
%%%%%%%%%%%%%%%%%%%%%%%%%%%%%%%%%%%%%%%
%  Slash character...
\def\slashchar#1{\setbox0=\hbox{$#1$}           % set a box for #1
   \dimen0=\wd0                                 % and get its size
   \setbox1=\hbox{/} \dimen1=\wd1               % get size of /
   \ifdim\dimen0>\dimen1                        % #1 is bigger
      \rlap{\hbox to \dimen0{\hfil/\hfil}}      % so center / in box
      #1                                        % and print #1
   \else                                        % / is bigger
      \rlap{\hbox to \dimen1{\hfil$#1$\hfil}}   % so center #1
      /                                         % and print /
   \fi}                                        %

%%EXAMPLE:  $\slashchar{E}$ or $\slashchar{E}_{t}$
\setcounter{tocdepth}{2}

%%%%%%%%%%%%%%%%%%%%%%%%%%%%%%%%%%%%%%%%%%%%%

{
\title{Neutrino mass constraints on $\beta$ decay}

\author{Takeyasu M. Ito}\affiliation{The Department of Physics and
  Astronomy, The University of Tennessee, Knoxville, TN 37996
\\and Physics Division, Oak Ridge National Laboratory, Oak Ridge, TN, 37831
}

\author{Gary Pr{\'e}zeau}\affiliation{Jet Propulsion
  Laboratory/California Institute of  Technology, 4800 Oak Grove Dr,
  Pasadena, CA 91109, USA}

\date{\today}

\begin{abstract}
Using the general connection between the upper limit on the neutrino
mass and the upper limits on certain types of non-Standard Model
interaction that can generate loop corrections to the neutrino mass,
we derive constraints on some non-Standard Model $d\rightarrow
ue^-\bar{\nu}$ interactions. When cast into limits on $n\rightarrow
pe^-\bar{\nu}$ coupling constants, our results yield constraints on
scalar and tensor weak interactions improved by more than an order of
magnitude over the current experimental limits.  When combined with
the existing limits, our results yield $|C_S/C_V|\alt 5\times
10^{-3}$, $|C'_S/C_V|\alt 5\times 10^{-3}$, $|C_T/C_A| \alt 1.2\times
10^{-2}$ and $|C'_T/C_A| \alt 1.2\times 10^{-2}$.  
\end{abstract}


\maketitle
}

%
% Introdcution
% neutron beta decay, its role in testing SM etc
% UCNA, abBA
Historically, nuclear $\beta$ decay has played an important role in
establishing the $V-A$ structure of the electroweak current of the
Standard Model (SM). More recently, precision studies of nuclear and
neutron $\beta$ decay have been used to test the SM and to search for
what may lie beyond it. $ft$-values and various angular correlations,
for example, have been measured on various nuclear species for small
deviations from what the $V-A$ model of weak interactions
predicts. These experiments have provided important constraints (for a
recent review, see Ref.~\cite{HER01}). With an increased intensity of
cold and ultracold neutrons becoming available, increasingly more
precise $\beta$-decay measurements with free neutrons may probe
physics beyond the SM. Neutron $\beta$-decay measurements have the
special advantage of being free from uncertainties due nuclear
structure corrections.  The UCNA experiment~\cite{UCNA} at the Los
Alamos National Laboratory, and the future abBA
experiment~\cite{ABBA}, planned for the Spallation Neutron Source at
the Oak Ridge National Laboratory, both aim at precision neutron
$\beta$-decay measurements that will provide stringent tests of the
SM.

On the other hand, various solar, atmospheric and reactor neutrino
experiments have provided clear evidence of neutrino oscillation,
hence establishing that not all the neutrinos are
massless~\cite{SK,SNO,Kamland}. In addition, the recent remarkable
progress in observational cosmology now allows us to study the
``Particle Physics'' of the early universe through precision measurements
of the anisotropy of the Cosmic Microwave Background (CMB). In fact,
the most stringent upper limit on the neutrino mass comes from combining
{\em WMAP}~\cite{SPE03} and  SDSS~\cite{Tegmark:2003ud} data.

The fact that the neutrino masses are so much smaller than the other
SM fermions -- at least six orders of magnitude -- together with the
fact that the lepton mixing matrix is strikingly different from the
quark mixing matrix, may be a window onto new physics. Accordingly,
the neutrino mass matrix has become a subject of intensive
experimental and theoretical research. At the same time, the search
for new physics through low-energy observables such as muon decay and
$\beta$ decay continues with increasing accuracy.  In view of this
situation, model-independent connections between the neutrino mass and
other low-energy observables would provide valuable guidance in the
search for physics beyond the SM.

% Prezeau et al
Recently, an important connection has been pointed out between the
neutrino mass and non-SM neutrino-matter interactions in
Ref.~\cite{PRE04}. That is, if there are non-SM neutrino-matter
interactions that involve both right-handed and left-handed neutrinos,
they should contribute to the neutrino mass. In Ref.~\cite{PRE04},
such contributions were evaluated using effective field theory; the
requirement that they be smaller than the current neutrino mass limits
resulted in non-trivial constraints on various muon decay parameters
and the branching ratio of the SM-forbidden
$\pi^0\rightarrow\nu\bar{\nu}$.

In this letter, we extend this treatment to $\beta$ decay, and obtain
order-of-magnitude constraints on some non-SM $n\rightarrow
pe^-\bar{\nu}$ interactions.  Although we closely follow the notation
of Ref.~\cite{HER01}, we generalize it to include the possibility of
total (and family) lepton-number violation.  Therefore, the most
general $d\rightarrow ue^-\bar{\nu}$ four-fermion interaction
involving both left-handed and right-handed neutrino states can be
written as
\begin{equation}
\label{eq:hamiltonian}
H_{\beta} = H_{V,A}+H_{S,P}+H_T,
\end{equation}
%where 
\begin{equation}
\label{eq:va}
H_{V,A} = 4\sum_{{\epsilon,\mu=\{\tl,\tr\}} \atop {l = e,\mu,\tau}}a_{\epsilon\mu}
\bar{e}\gamma^{\lambda}P_{\epsilon}\nu^{(c)}_l 
\bar{u}\gamma_{\lambda}P_{\mu}d 
+ \text{h.c.},
\end{equation}
\begin{equation}
\label{eq:sp}
H_{S,P} = 4\sum_{{\epsilon,\mu=\{\tl,\tr\}} \atop {l = e,\mu,\tau}} 
A_{\epsilon\mu}
\bar{e}P_{\epsilon}\nu^{(c)}_l 
\bar{u}P_{\mu}d
+ \text{h.c.},
\end{equation}
\begin{equation}
\label{eq:t}
H_{T} = 
4\sum_{{\epsilon=\{\tl,\tr\}} \atop {l = e,\mu,\tau}}{\alpha}_{\epsilon\mu}
\bar{e}\frac{\sigma^{\alpha\beta}}{\sqrt{2}}
P_{\epsilon}\nu^{(c)}_l 
\bar{u}\frac{\sigma_{\alpha\beta}}{\sqrt{2}}
P_{\epsilon}d 
+ \text{h.c.}
\end{equation}
where, $P_{\tl,\tr}=(1\mp\gamma^5)/2$ is the chirality projection
operator, $\epsilon$ and $\mu$ denote the chiralities of the neutrino
and the $d$ quark, respectively, the superscript $(c)$ on the neutrino
indicates charge-conjugation for the case where the operators violate
total lepton-number conservation, while the subscript $l$ takes into
account the possibility of family lepton-number violation.  In the SM,
$a_{\tl\tl}=a_{\tl\tl}^{\text SM}\equiv g^2 V_{ud}/8m_W^2$ and all the
other coupling constants $a_{\epsilon\mu}$, $A_{\epsilon\mu}$, and
$\alpha_{\epsilon\mu}$ are 0. The $d\rightarrow ue^-\bar{\nu}$
coupling constants $a_{\epsilon\mu}$, $A_{\epsilon\mu}$ and
$\alpha_{\epsilon\mu}$ can be related to more conventionally used
$n\rightarrow pe^-\bar{\nu}$ coupling constants $C_{i}$ and $C'_{i}$
($i=\{V,A,S,T\}$)~\cite{JAC57}, as shown in the Appendix. In the SM,
$C'_V=-C_V$, and $C'_A=-C_A$.

\begin{figure}[tb]
\resizebox{8 cm}{!}{\includegraphics*[35,520][520,720]{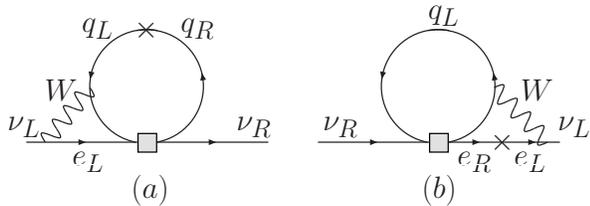}}
\caption{Two-loop contributions to the neutrino mass generated by
  chirality-changing non-SM $d\rightarrow ue^-\bar{\nu}$ operators.
  The $\times$ denote mass insertions.  Fig.~(a) constrains
  $A_{\tr\tr},~ A_{\tr\tl},~\alpha_{\tr\tr}$ and involves a quark mass
  insertion $m_q\approx 4$~MeV while Fig.~(b) constrains $a_{\tr\tl}$
  and requires an electron mass insertion, $m_e=0.511$~MeV.
} \label{fig:beta2loop}
\end{figure}

%Connection between neutrino mass and non-SM e\nu ud interaction
As discussed in Ref.~\cite{PRE04}, certain types of non-SM
interactions generate contributions to neutrino mass through loop
effects. In this case,
the $A_{\tr\tr}$-, $A_{\tr\tl}$-, and $\alpha_{\tr\tr}$-type
interactions contribute to the neutrino mass through the diagram shown
in Fig.~\ref{fig:beta2loop}a, while the $a_{\tr\tl}$-type interaction
contributes through the diagram in
Fig.~\ref{fig:beta2loop}b~\footnote{The $a_{\tr\tr}$-type
  interaction also contributes to the neutrino mass. We do not
  consider it here, however, since a stringent constraint on
  $a_{\tr\tr}$ cannot be arrived at by the method discussed in this
  paper.}.

Following Ref.~\cite{PRE04}, we evaluate the leading log contributions
to the neutrino mass from the diagrams in Fig.~\ref{fig:beta2loop}
using dimensional regularization. The result is
%
%%%%%%%%%%%%%%%%%%%%%%%%%%%%%%%%%%%%%%%%%%%%%%%%%%%%%%%%%%%%%%%%%%%%
\beqa \label{eq:delta-m}
\delta m_\nu \approx
g^2 \nc \gf \bar{a} \frac{m_f M_W^2}{(4 \pi)^4}
\left( \ln{\mu^2\over M_W^2} \right)^2 ~,
\label{eq:deltam}
\eeqa
%%%%%%%%%%%%%%%%%%%%%%%%%%%%%%%%%%%%%%%%%%%%%%%%%%%%%%%%%%%%%%%%%%%%
where $N_c$ is the number of color degrees of freedom, $G_F$ is the
Fermi constant $G_F=1.166\times 10^{-5}$~GeV$^{-2}$, $m_f$ is the mass
of the fermion for the mass insertion, $g\cong 0.64$ is the SU(2)
gauge coupling constant, and $\mu$ is the renormalization
scale discussed below. $\bar{a}=\{\bar{A}_{\tr\tr}, \bar{A}_{\tr\tl},
\bar{\alpha}_{\tr\tr}, \bar{a}_{\tr\tl}\}$ is the ``normalized''
$d\rightarrow ue^-\bar{\nu}$ coupling, with
$\bar{A}_{\tr\tr}=A_{\tr\tr}/a_{\tl\tl}^{\text SM}$, etc.  $\bar{a}$
in principle can have four indices denoting the flavors of incoming
and outgoing leptons and quarks. We suppress the indices here,
however, for simplicity. We ignored the factors of half or two in the
loop integrals associated with different Dirac structures, since we
are interested in the orders of magnitude.

%Remornalization
The value of $\mu$ should exceed the mass of the heaviest particle
included in the effective field theory -- in our case $M_W$ -- while
at the same time take into account the scale at which the onset of new
physics might be expected. We choose the renormalization scale to be
around 1 TeV, a scale often associated with physics beyond the SM in
many particle physics models. Since $\mu$ appears in a logarithm, our
conclusions do not depend strongly on its precise value.
%%%% Begin new text from Gary
Note that for values of $\mu$ below $\sim 100$~GeV, the decoupling
theorem kicks in and large logarithms become suppressed by inverse
powers of the weak boson mass.  This can be easily seen in the
physical renormalization scheme where loop corrections are required to
vanish at the renormalization point.  In that case, the logarithms
will have the generic form $\ln|(\mu^2 - M_W^2)/(Q^2 - M_W^2)|$.  At
$Q^2=\mu^2$, this vanishes as required.  For small values of $Q^2$,
large logarithms of the form $\ln(\mu^2/M_W^2)$ appear only for values
of $\mu^2>>M_W^2$.  For $\mu^2<<M_W^2$, the large logarithms are
suppressed by inverse powers of $M_W^2$.  It follows that arbitrarily
small values of $\mu$ cannot be inserted in Eq.~(\ref{eq:delta-m})
which was evaluated for $\mu > M_W$.
%%%% End new text from Gary
A more detailed discussion of the dependence of
$\delta \mn$ on the renormalization scale is given in
Ref.~\cite{PRE04}.

%%% Begin New Text from Gary
The interactions of the Hamiltonian in Eq.~(\ref{eq:hamiltonian}) are
not gauge-invariant under SU$(2)_{\text{L}}\times
\text{U}_{\text{Y}}$, although they can arise from gauge-invariant
models, e.g., the left-right symmetric and lepto-quark models
\cite{HER01}.  Thus, the evaluation of the diagrams of
Fig.~\ref{fig:beta2loop} above the weak scale is not strictly
orthodox; in principle, one should evaluate contributions to the
neutrino mass within each gauge-invariant model to constrain the
relevant parameters appearing in the particular model.  The advantage
of the current approach stems from the emphasis on the physics that
all extensions of the SM that generate the chirality-flipping
interactions of Eq.~(\ref{eq:hamiltonian}) share: they all contain
operators that contribute to the neutrino mass resulting in
constraints on a class of model parameters.  The rough estimates given
in Eq.~(\ref{eq:deltam}) make this novel point in an essentially
model-independent way; these estimates are expected to be
representative of the constraints imposed on the parameters due to the
smallness of the neutrino mass, although this should be checked in
specific models.  Indeed, the neutrino mass is not required to vanish
by gauge-invariance.  Therefore, there is no reason to expect
cancellations between the diagrams that contribute to $m_\nu$ in
gauge-invariant models.
%%% End New Text from Gary

%Place limits on e\nu ud interaction
We use Eq.~(\ref{eq:delta-m}) to constrain $\bar{a}$ by requiring
$\delta m_\nu < m_{\nu}$ where $m_{\nu}$ is the physical neutrino
mass. As in Ref.~\cite{PRE04} we adopt the upper limit of 0.71~eV on
the sum of the neutrino masses from Ref.~\cite{SPE03}. This implies
the limit $m_{\nu} < 0.23$~eV for individual neutrino masses when
neutrino oscillation constraints are included~\cite{PDG}. 
%%%%Note that since we use the neutrino mass limits obtained
%%%%from the CMB measurements that are only sensitive to the masses of the
%%%%light neutrinos, the constraints derived here are only applicable to
%%%%processes that involve light neutrinos.

For Fig.~\ref{fig:beta2loop}~(a), from Eq.~(\ref{eq:delta-m}) with
$m_f=m_{u,d} \approx 4$~MeV, we obtain $\bar{A}_{\tr\tr} < 10^{-3}$,
$\bar{A}_{\tr\tl} < 10^{-3}$, and $\bar{\alpha}_{\tr\tr} <
10^{-3}$. For Fig.~\ref{fig:beta2loop}~(b), with $m_f=m_e$, we obtain
$\bar{a}_{\tr\tl}<10^{-2}$. If one takes a neutrino mass limit $<
0.04$~eV~\cite{HAN03} possibly reached by the future Planck
mission~\cite{PLANCK}, one obtains $\bar{A}_{\tr\tr} < 10^{-4}$,
$\bar{A}_{\tr\tl} < 10^{-4}$, $\bar{\alpha}_{\tr\tr} < 10^{-4}$, and
$\bar{a}_{\tr\tl}<10^{-3}$. The obtained constraints on $\bar{a}$ are
summarized in Table.~\ref{tab:results} together with experimental
limits, and constraints on quantities derived from $\bar{a}$, which we
discuss below.

%Discussion of the results
Our limit $\bar{a}_{\tr\tl}<10^{-2}$ is comparable to the present
experimental limit from $\beta$ decay $\bar{a}_{\tr\tl}<3.7\times
10^{-2}$~\cite{DEU99} (see also Ref.~\cite{HER01}).  
%Our limit is
%improved by an order of magnitude if the Planck mission constrains the
%neutrino mass to $m_{\nu} < 0.04$~eV.

As seen from the equations in the Appendix, $A_{\tr\tr}$, $A_{\tr\tl}$,
and $\alpha_{\tr\tl}$ are related to the $n\rightarrow pe^-\bar{\nu}$
coupling constants as follows:
\begin{equation}
2g_S (A_{\tr\tr} + A_{\tr\tl})=C_S + C'_S,
\end{equation}
and
\begin{equation}
4g_T \alpha_{\tr\tr} = C_T + C'_T.
\end{equation}
Our results yield order-of-magnitude constraints of
\begin{equation}
|\tilde{C}_S + \tilde{C}'_S| \alt 10^{-3},
\end{equation}
and
\begin{equation}
|\tilde{C}_T + \tilde{C}'_T| \alt 10^{-2},
\end{equation}
where $\tilde{C}_S = C_S/C_V$, $\tilde{C}'_S = C'_S/C_V$, $\tilde{C}_T
= C_T/C_A$, and $\tilde{C}'_T = C'_T/C_A$. We used $C_V \cong g_V
a_{\tl\tl}^{\text SM}$, $C_A \cong g_A a_{\tl\tl}^{\text SM}$, and
$0.25 \alt g_S \alt 1$ and $0.6 \alt g_T \alt 2.3$~\cite{HER01,ADL75}.

\begin{figure}[tb]
\resizebox{14pc}{!}{\includegraphics{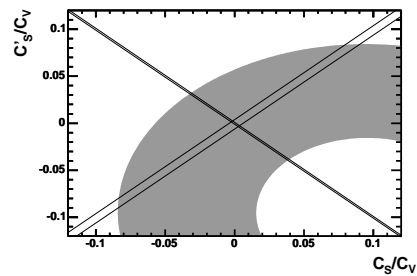}}
\caption{Constraints on $\tilde{C}_S=C_S/C_V$ and
  $\tilde{C}'_S=C_S/C_V$. The narrow diagonal band at $-45^o$ is from
  this work. The annulus gray is a 95\% C.L. limit from
  Ref.~\cite{ADE99}. The diagonal band at $45^o$ is a 90\% C.L. limit
  from Ref.~\cite{ORM89}.}
\label{fig:scalar}
\end{figure}

The current experimental limits on $C_S$ and $C'_S$ come from the
$e^+$-$\nu$ correlation in $^{32}$Ar~$\beta$ decay~\cite{ADE99} and
the $ft$ values of super-allowed $\beta$ decays~\cite{ORM89} (An
updated analysis gives a similar limit~\cite{GAR98}).  A combined
analysis of data from Refs.~\cite{ADE99} and \cite{ORM89} gives a
one-standard deviation bound of $|\tilde{C}_s|^2 \leq 3.6\times
10^{-3}$ and $|\tilde{C}'_s|^2 \leq 3.6\times 10^{-3}$~\cite{ADE99},
which implies $|\tilde{C}_S + \tilde{C}'_S| \alt 10^{-1}$. Our
constraints are more stringent by two orders of magnitude and are
compared with the existing limits in Fig.~\ref{fig:scalar} where it is
seen that they are complimentary to the existing limits. Combining our
results with the existing limits yields $|\tilde{C}_S|\alt 5\times
10^{-3}$ and $|\tilde{C}'_S|\alt 5\times 10^{-3}$.

\begin{figure}[tb]
\resizebox{14pc}{!}{\includegraphics{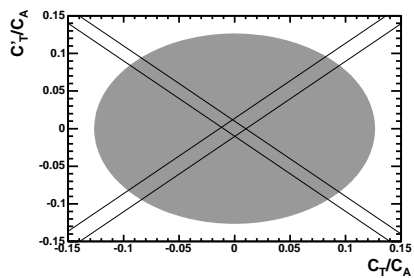}}
\caption{Constraints on $\tilde{C}_T=C_T/C_A$ and
  $\tilde{C}'_T=C'_T/C_A$. The diagonal band at $-45^o$ is from this
  work. The gray circle is a 68\% C.L. limit from
  Ref.~\cite{GLU89}. The diagonal band at $45^o$ is a 90\% C.L. limit
  from Ref.~\cite{CAR91}.}
\label{fig:tensor}
\end{figure}

For the tensor interaction, the present experimental limit is provided
by $^6$He $\beta$ decay~\cite{GLU89} and the positron polarization of
$^{14}$O and $^{10}$C $\beta$ decay~\cite{CAR91}. Ref.~\cite{GLU89}
quotes $(|C_T|^2 +|C'_T|^2)/(|C_A|^2 + |C'_A|^2) < 0.8\%$ (68\% C.L.),
which implies $|\tilde{C}_T+\tilde{C}'_T| \alt 1.6\times 10^{-1}$. Our
results provide a constraint improved by an order of magnitude. Our
results are shown in Fig.~\ref{fig:tensor} together with the current
experimental limits~\footnote{If one allowed $b\neq 0$ in analyzing
  the $^6$He $\beta$ decay data, one would obtain an annulus for the
  allowed region as in the $^{32}$Ar case. However, this would not
  affect the main conclusion of this paper.}.  When combined with the
existing limits, our results yield $|\tilde{C}_T| \alt 1.2\times 10^{-2}$
and $|\tilde{C}'_T| \alt 1.2\times 10^{-2}$.

\begin{table}
\caption{Constraints on $d\rightarrow ue^-\bar{\nu}$ coupling
  constants $\bar{a}=\{\bar{A}_{\tr\tr}, \bar{A}_{\tr\tl},
  \bar{\alpha}_{\tr\tr}, \bar{a}_{\tr\tl}\}$ obtained from this study
  (top) and constraints on quantities derived from
  $\bar{a}$ (bottom) together with current
  experimental limits.}
\label{tab:results}
\begin{ruledtabular}
\begin{tabular}{ccc}
$\bar{a}$ & Current limits & Limits from this study \\ \hline
$|\bar{A}_{\tr\tr}+\bar{A}_{\tr\tl}|$& $\sim 0.1$ &$\sim 2\times 10^{-3}$\\
$|\bar{\alpha}_{\tr\tr}|$ & $8\times 10^{-2}$ (68\% c.l.) & $\sim 10^{-3}$ \\
$|\bar{a}_{\tr\tl}|$ & $3.7\times 10^{-2}$ (90\% c.l.) & $\sim 10^{-2}$\\
\end{tabular}
\vspace{2mm}
\begin{tabular}{ccc}
$n\rightarrow pe^-\bar{\nu}$ Coupling & 
Current limits & Limits from this study \\ \hline
$|\tilde{C}_S+\tilde{C}'_S|$& $\sim 0.1$ & $\sim 10^{-3}$ \\
$|\tilde{C}_T+\tilde{C}'_T|$& $1.6\times 10^{-1}$ & $\sim 10^{-2}$\\
\end{tabular}
\end{ruledtabular}
\end{table}


The initial goal of the UCNA experiment is to measure the
$\beta$ asymmetry parameter $A$ using ultracold neutrons at the 0.2\%
level. With an 
implementation of additional detectors, the UCNA experiment also aims to
measure the $e^--\bar{\nu}_e$ angular coefficient $a$, the
$\bar{\nu}_e$ asymmetry parameter $B$, and the Fierz interference
coefficient $b$,  with the following accuracies:
$\delta_a/a \leq 3\times 10^{-3}$, $\delta_A/A \leq 10^{-3}$,
$\delta_B/B \leq 10^{-3}$, and $\delta_b \leq 2\times 10^{-3}$.  The
abBA experiment aims to measure the same quantities with similar
accuracies using a pulsed cold neutron beam.

With such precision, it is likely that these free neutron experiments
will constrain the scalar interactions by a factor of two better than
the current experimental limits~\cite{GAR04}.  In general
$\beta$-decay experiments are mostly sensitive to $C_S - C'_S$ and
$C_T - C'_T$ through $b$ measurements and $|C_S|^2 +|C'_S|^2$ and
$|C_T|^2 + |C'_T|^2$ through $a$ measurements, while our results
provide constraints on $C_S + C'_S$ and $C_T + C'_T$, therefore making
our analysis and these experiments complimentary.

The results presented here are quite general and are valid for both
Dirac and Majorana neutrinos.  As discussed in Ref.~\cite{PRE04},
there may be cases where the neutrino mass constraints are beaten, for
example in the presence of finely tuned cancellations between various
Feynman graphs.  Also, we do not take into account effects stemming
from neutrino mixing with heavy mass eigenstates since in most models
they are much heavier than the energy released in $\beta$-decay and
their emission is kinematically forbidden. If the tritium beta decay
limit on neutrino mass ($m_{\nu_e} \alt 3$~eV)~\cite{PDG} is used
instead of the CMB limit, our analysis still yields limits an order of
magnitude more stringent than the current experimental limits for the
scalar couplings. We also point out that the KATRIN experiment will
have sensitivity down to $m_{\nu_e} \sim 0.2$~eV~\cite{OSI01}.

%%%Also, the applicability of our results is limited
%%%because of the use of the neutrino mass limits obtained from the CMB
%%%measurements. Since the CMB measurements are only sensitive to light
%%%neutrinos, our constraints on non-SM $d\rightarrow ue^-\bar{\nu}$
%%%couplings only apply to processes that involve light
%%%neutrinos. However, non-SM $n\rightarrow pe^-\bar{\nu}$ interactions
%%%manifest themselves in $\beta$-decay observables only when they
%%%involve light neutrinos. Therefore, as long as our results are applied
%%%to $\beta$-decay observables, the use of the CMB neutrino mass limits
%%%does not limit the applicability of our results. 
%If future
%$\beta$-decay experiments yielded results that contradict our
%constraints, it would certainly cause an interesting situation to
%arise as it would have an implication for the neutrino mass, or would
%indicate fine-tuned cancellations of the radiative corrections to the
%neutrino mass.

In conclusion, we have calculated the leading contributions to the
neutrino mass generated by non-SM $d\rightarrow ue^-\bar{\nu}$
interactions that involve right-handed neutrinos. Using the current
upper limits on the neutrino mass obtained from CMB measurements, we
derived order-of-magnitude constraints on the corresponding coupling
constants.  When cast into the effective $n\rightarrow pe^-\bar{\nu}$
coupling constants, our results improve over the current experimental
constraints on the scalar and tensor coupling constants by more than
an order of magnitude. When combined with the existing limits, our
results yield $|C_S/C_V|\alt 5\times 10^{-3}$, $|C'_S/C_V|\alt 5\times
10^{-3}$, $|C_T/C_A| \alt 1.2\times 10^{-2}$ and $|C'_T/C_A| \alt
1.2\times 10^{-2}$.

One of the authors (T.~M.~I) thanks A.~Garcia for valuable 
discussions.

\appendix*
\section{Appendix}
Here also, we follow the notation of Ref.~\cite{HER01}. Neglecting the
induced from factors, the effective $n\rightarrow pe^-\bar{\nu}$
interaction is given by
\begin{equation}
H_{\beta}^{(N)} \sim H_{V,A}^{(N)} + H_S^{(N)} + H_T^{(N)},
\end{equation}
where
\begin{eqnarray}
H_{V,A}^{(N)} & = &
\bar{e}\gamma_{\lambda}(C_V+C'_V\gamma_5)\nu_e
\bar{p}\gamma^{\lambda}n \nonumber\\
& + &
\bar{e}\gamma_{\lambda}\gamma_5(C_A+C'_A\gamma_5)\nu_e
\bar{p}\gamma^{\lambda}n + \text{h.c.}, 
\end{eqnarray}
\begin{equation}
H_S^{(N)} = \bar{e}(C_S + C'_S\gamma_5)\nu_e\bar{p}n+\text{h.c.},
\end{equation}
\begin{equation}
H_T^{(N)} = \bar{e}\frac{\sigma^{\lambda\mu}}{\sqrt{2}}
(C_T +C'_T\gamma_5)\nu_e\bar{p}\frac{\sigma_{\lambda\mu}}{\sqrt{2}}n
+\text{h.c.}
\end{equation}
The $n\rightarrow pe^-\bar{\nu}$ coupling constants $C_{i}$ and
$C'_{i}$ ($i=\{V,A,S,T\}$) are related to the $d\rightarrow
ue^-\bar{\nu}$ coupling constants $a_{\epsilon\mu}$,
$A_{\epsilon\mu}$ and $\alpha_{\epsilon\mu}$
($\epsilon\mu=\tr,\tl$) as follows:
\begin{equation}
C_V,\; C'_V = g_V(\pm a_{\tl\tl}\pm a_{\tl\tr}+a_{\tr\tr}+a_{\tr\tl}),
\end{equation}
\begin{equation}
C_A,\; C'_A = g_A(\pm a_{\tl\tl}\mp a_{\tl\tr}+a_{\tr\tr}-a_{\tr\tl}),
\end{equation}
\begin{equation}
C_S,\; C'_S = g_S(\pm A_{\tl\tl}\pm A_{\tl\tr}+A_{\tr\tr}+A_{\tr\tl}),
\end{equation}
\begin{equation}
C_T,\; C'_T = 2g_T(\pm \alpha_{\tl\tl}+\alpha_{\tr\tr}),
\end{equation}
where the upper and lower signs are for $C_i$ and $C'_i$,
respectively. The constants $g_i=g_i(0)$ are the $q^2\rightarrow 0$
values of the nucleon form factors defined by
\begin{equation}
<p|\bar{u}\Gamma_id|n>=g_i(q^2)\bar{p}\Gamma_in,
\end{equation}
where $i=\{V,A,S,T\}$, and $\Gamma_V=\gamma_{\lambda}$,
$\Gamma_A=\gamma_{\lambda}\gamma_5$, $\Gamma_S=1$, and
$\Gamma_P=\gamma_5$. CVC predicts $g_V=1$. $g_A=1.2695(29)$
~\cite{PDG} (our definition of $g_A$ differs from that adopted by
Particle Data Group by the sign).

\end{document}